\pdfoutput=1 
\documentclass[a4paper, 11pt]{article}
\PassOptionsToPackage{table}{xcolor}

\usepackage{jheppub} 

\usepackage[T1]{fontenc} 
\usepackage{tangocolors}
\usepackage{enumitem}
\usepackage{tikz}
\usetikzlibrary{arrows, decorations,backgrounds, patterns}
\usetikzlibrary{decorations.pathreplacing ,decorations.markings}
\usepackage{amssymb}
\usetikzlibrary{decorations.pathmorphing,backgrounds,shapes,arrows,shadows}
\usetikzlibrary{patterns.meta}
\usetikzlibrary{patterns,decorations.pathmorphing}
\tikzset{
    snake it/.style={decorate, decoration=snake}
}
\usepackage{pgfplots}
\pgfplotsset{compat=1.11}
\usepgfplotslibrary{fillbetween}
\usetikzlibrary{intersections}
\pgfdeclarelayer{bg}
\pgfsetlayers{bg,main}
\tikzset{zigzag/.style={decorate,decoration=zigzag}}
\tikzset{snake it/.style={decorate, decoration=snake}}
\makeatletter
\def\@hex@@Hex#1%
 {\if a#1A\else \if b#1B\else \if c#1C\else \if d#1D\else
  \if e#1E\else \if f#1F\else #1\fi\fi\fi\fi\fi\fi \@hex@Hex}
\makeatother

\usepackage[all]{xy}

\makeatletter  
\gdef\@fpheader{}  
\makeatother 
\usepackage[percent]{overpic}
\usepackage{slashed}
\usepackage{wrapfig}
\usepackage{tabu}
\usepackage{diagbox}
\usepackage{mathrsfs,amsmath,amssymb,amsthm,amsfonts,tikz,graphicx,accents,hyperref, color}
\usepackage{dsfont,epiolmec, latexsym, stmaryrd, comment}
\usepackage{slashed,ccaption}
\usepackage{mathrsfs, calligra}
\usepackage{leftidx}
\usepackage{import}
\usepackage{multirow}
\usepackage{amsfonts}
\usepackage{pifont}
\usepackage{tabularx}
\usepackage{cancel}
\usepackage[utf8]{inputenc}
\usetikzlibrary{intersections,calc}
\usepackage{ifthen}
\usepackage{amsmath}
\usepackage{cancel}
\usepackage{caption} 
\usepackage{subcaption}

\usepackage{ragged2e}

\usetikzlibrary{patterns.meta}
\usepackage{array}
\hypersetup{ linktoc=all,
    colorlinks, linkcolor={palatinateblue},
    citecolor={brightpink}, urlcolor={amaranth}
}

\graphicspath{{Images/}}

\def\sideremark#1{\ifvmode\leavevmode\fi\vadjust{\vbox to0pt{\vss
 \hbox to 0pt{\hskip\hsize\hskip1em
 \vbox{\hsize2cm\tiny\raggedright\pretolerance10000
 \noindent #1\hfill}\hss}\vbox to8pt{\vfil}\vss}}}%
\DeclareSymbolFont{extraup}{U}{zavm}{m}{n}
\DeclareMathSymbol{\varheart}{\mathalpha}{extraup}{86}
\DeclareMathSymbol{\vardiamond}{\mathalpha}{extraup}{87}
\makeatletter
\renewcommand*{\@fnsymbol}[1]{\ensuremath{\ifcase#1\or \clubsuit \or \vardiamond \or \varheart\or
    \spadesuit\or \mathparagraph\or \|\or **\or \dagger\dagger
    \or \ddagger\ddagger \else\@ctrerr\fi}}
\makeatother

\definecolor{rosy}{RGB}{230,235,252}
\definecolor{myframetitle}{RGB}{90,89,170}
\definecolor{myblocktitle}{RGB}{140,185,249}
\definecolor{mytitle}{RGB}{10,80,26}

\definecolor{darkgreen}{RGB}{27,130,45}
\definecolor{darkblue}{rgb}{0,0,0.3}
\definecolor{darkred}{rgb}{0.7,0,0}

\definecolor{light gray}{RGB}{220,220,220}
\definecolor{dark purple}{RGB}{108,0,217}
\definecolor{pink}{RGB}{190,20,100}
\definecolor{orang}{RGB}{193,63,0}
\definecolor{green}{RGB}{11,98,17}
\definecolor{darkpink}{RGB}{153,0,76}
\definecolor{bluegreen}{RGB}{0,102,102}
\definecolor{greenlagan}{RGB}{0,102,0}
\definecolor{redgreen}{RGB}{102,102,0}
\definecolor{Redgreen}{RGB}{153,76,0}
\definecolor{vividviolet}{rgb}{0.62, 0.0, 1.0}
\definecolor{amaranth}{rgb}{0.9, 0.17, 0.31}
\definecolor{palatinateblue}{rgb}{0.15, 0.23, 0.89}
\definecolor{brightpink}{rgb}{1.0, 0.0, 0.5}
\definecolor{cornflowerblue}{rgb}{0.39, 0.58, 0.93}
\definecolor{deepcarminepink}{rgb}{0.94, 0.19, 0.22}
\definecolor{radicalred}{rgb}{1.0, 0.21, 0.37}

\usepackage[most]{tcolorbox}

\tcbset{highlight math style={left=02mm,right=02mm,top=02mm,bottom=02mm}} 
\usepackage{empheq}

\newcommand\inbox[1]{\tcbset{fonttitle=\scriptsize} \tcboxmath[colback=white,colframe=black!70]{#1}}
\usetikzlibrary{calc} 

\DeclareFontFamily{OT1}{rsfs}{}

\DeclareFontShape{OT1}{rsfs}{m}{n}{ <-7> rsfs5 <7-10> rsfs7 <10->rsfs10}{} 

\DeclareMathAlphabet{\mycal}{OT1}{rsfs}{m}{n}

\newcommand{\be}{\begin{equation}}
\newcommand{\ee}{\end{equation}}
\newcommand{\bea}{\begin{eqnarray}}
\newcommand{\eea}{\end{eqnarray}}

\newcommand{\Ottbar}{{T\hspace*{-1mm}{T}}}
\newcommand{\ttbar}{{\mathcal{T}\hspace*{-1mm}{\mathcal{T}}}}

\makeatletter \@addtoreset{equation}{section}

\begin{document}


\newcommand{\mytitle}{{\textbf{\centerline{\LARGE{Freelance Holography}}\\ }}}

\title{{\mytitle}}
\author[]{Vahid~Taghiloo}
\affiliation{School of Physics, Institute for Research in Fundamental
Sciences (IPM),\\ P.O.Box 19395-5531, Tehran, Iran}
\affiliation{Department of Physics, Institute for Advanced Studies in Basic Sciences (IASBS),\\ 
P.O. Box 45137-66731, Zanjan, Iran}

\emailAdd{
v.taghiloo@iasbs.ac.ir
}

\abstract{
In this paper, we introduce the Freelance Holography Program, an extension of the AdS/CFT correspondence within the saddle-point approximation that opens several novel directions. This framework generalizes holography beyond the asymptotic AdS boundary, allowing it to be formulated on arbitrary timelike hypersurfaces in the bulk. Moreover, it accommodates arbitrary boundary conditions for bulk fields, moving beyond the standard Dirichlet prescription. As part of this development, we construct a one-parameter family of renormalized boundary conditions that, unlike conventional choices in the literature, lead to a finite on-shell action. We also explore intriguing consequences of the framework, including the emergence of induced gravity and the flow of boundary conditions under holographic renormalization.
}


\dedicated{
\begin{center}
    This letter won the {Second Prize} in the ``2025 Prizes for Letters on Holography,'' officially awarded by the \textit{Journal of Holography Applications in Physics} (JHAP). 
It is accepted for publication in JHAP. See the \href{https://jhap.du.ac.ir/news?newsCode=23}{official page}.
\end{center}
}

\maketitle

\section{Introduction}
Motivated by the Bekenstein–Hawking formula for black hole entropy \cite{Bekenstein:1973ur, Hawking:1974sw}, ’t Hooft and Susskind proposed the \textit{holographic principle}: a theory of quantum gravity in $d{+}1$ dimensions can be described by a non-gravitational quantum field theory in $d$ dimensions \cite{tHooft:1993dmi, Susskind:1993aa}. Its best-understood realization is the \textit{AdS/CFT correspondence}, which posits that quantum gravity in asymptotically anti-de Sitter (AdS) spacetime is dual to a conformal field theory (CFT) on the AdS boundary \cite{Maldacena:1998re, Witten:1998qj, Gubser:1998bc}.

Despite its wide applicability, the AdS/CFT correspondence has several notable limitations: 1) It applies specifically to \textit{asymptotically AdS spacetimes}. 2) As we will review, the duality requires \textit{Dirichlet} boundary conditions for bulk fields on the asymptotic timelike boundary. 3) The dual non-gravitational theory is defined only on the \textit{asymptotic boundary} of AdS. Various efforts have addressed the first issue by extending holography to spacetimes with alternative asymptotics—for instance, codimension-one and codimension-two holography in asymptotically flat spacetimes \cite{Bagchi:2025vri, Strominger:2017zoo}, and constructions in asymptotically de Sitter (dS) spacetimes using different frameworks \cite{Strominger:2001pn, Susskind:2021omt}.

In this work, we focus on limitations (2) and (3). We develop a framework that extends holography in the \textit{saddle-point approximation}, where the bulk theory is classical and the boundary theory is in the large-$N$ limit. This framework accommodates \textit{arbitrary} boundary conditions for bulk fields at the asymptotic boundary of AdS. More significantly, we go beyond the standard setup by introducing \textit{finite-cutoff holography}, which proposes a duality between a finite region of AdS—bounded by a timelike cutoff surface—and a boundary theory defined on that finite-radius hypersurface. As a further step, we extend this to allow arbitrary boundary conditions on the cutoff surface. We call this generalized approach \textit{Freelance Holography}, as it frees the AdS/CFT correspondence from the constraints of fixed boundary conditions and a fixed asymptotic boundary, broadening the holographic paradigm.

{A central motivation for developing holography with arbitrary boundary conditions, particularly at a finite radial position, stems from the well-posedness of gravitational boundary value problems. While Dirichlet boundary conditions at the asymptotic AdS boundary yield a well-posed problem, it has been shown that imposing Dirichlet boundary conditions at a \textit{finite} radial position does \textit{not} generally lead to a well-posed boundary value problem \cite{Anderson:2006lqb, Witten:2018lgb}. However, there is evidence that other classes of boundary conditions, such as the seminal \textit{conformal} boundary condition \cite{Anderson:2006lqb, Witten:2018lgb}, do result in well-posed dynamics. This observation implies that constructing finite-cutoff holography based solely on Dirichlet boundary conditions is fundamentally problematic. Consequently, a more general formulation of finite-cutoff holography—one that allows for \textit{non-Dirichlet} boundary conditions—is not only natural but also necessary.}

{
The foundational ideas and technical framework of freelance holography were first developed in \cite{Parvizi:2025shq, Parvizi:2025wsg}, building on the covariant phase space formalism (CPSF) \cite{Lee:1990nz, Iyer:1994ys, Wald:1999wa}. As a striking application, we have recently shown that gravity is not fundamental but rather emerges through the renormalization group flow of the boundary theory \cite{Adami:2025pqr}.}

{In this letter, we revisit the program from a fresh perspective, deliberately avoiding CPSF to reduce technical overhead and present new proofs, making the framework more accessible to a broader range of researchers. We emphasize the constrained nature of the boundary deformation flow equation and highlight how boundary gravity emerges naturally through this flow in arbitrary spacetime dimensions. Furthermore, we clarify the distinction between renormalized and unrenormalized boundary conditions, discussing their fundamental differences. Together, these developments provide a streamlined and deeper presentation of freelance holography and its underlying principles.
}

\section{Review on gauge/gravity correspondence}
In this section, we briefly review the AdS/CFT duality \cite{Maldacena:1998re, Witten:1998qj, Gubser:1998bc} in the saddle-point approximation, where the bulk is a weakly coupled classical theory and the boundary theory is in the large-$N$ limit. In this regime, the correspondence reduces to the equality of the on-shell actions in bulk and boundary theories
\begin{equation}\label{AdS/CFT}
{S}^{\text{\tiny{D}}}_{\text{bdry}}[\mathcal{J};\Sigma] = S^{\text{\tiny{D}}}_{\text{bulk}}[\mathcal{J}; \mathcal{M}]\,.
\end{equation}
The right-hand side represents the on-shell action of $(d{+}1)$-dimensional gravity in asymptotically AdS spacetime
\begin{equation}
S^{\text{\tiny{D}}}_{\text{bulk}}[J; \mathcal{M}] = \int_{\mathcal{M}} \mathcal{L}^{\text{\tiny{D}}}_{\text{bulk}}\big|_{\text{on-shell}}\, .
\end{equation}
Here, $\mathcal{M}$ is the bulk spacetime, $J$ the bulk fields, and the superscript “D” indicates Dirichlet boundary conditions on the AdS boundary $\Sigma \equiv \partial \mathcal{M}$
\begin{equation}\label{D-bc}
\delta J(r,x^a)\big|_{\Sigma}=0\, , \qquad J(r,x^a)\big|_{\Sigma} = r_{\infty}^{d-\Delta} \mathcal{J}(x^a)\,,
\end{equation}
Here, $r_\infty$ denotes the asymptotic limit $r \to \infty$, corresponding to the AdS timelike boundary $\Sigma$ \footnote{Bulk coordinates are $x^{\mu} = {r, x^a}$, with $r$ the radial direction and $x^a$ the boundary coordinates ($a = 0, \dots, d{-}1$).}. The parameter $\Delta$ is the scaling dimension of the momentum conjugate to $J$. $\mathcal{J}(x^a)$ depends only on boundary coordinates and is non-dynamical due to the Dirichlet condition.

We now turn to the left-hand side of the AdS/CFT dictionary \eqref{AdS/CFT}
\begin{equation}\label{bdry-theory-infty}
{S}^{\text{\tiny{D}}}_{\text{bdry}}[\mathcal{J}; \Sigma] = S_{\text{\tiny{CFT}}}[\phi^*] + \int_{\Sigma} \sqrt{-\gamma}\, \mathcal{J}\, \mathcal{O}[\phi^*]\,.
\end{equation}
In this expression, $S_{\text{\tiny{CFT}}}$ is the action of a conformal field theory on the AdS boundary, with dynamical variables $\phi(x^a)$. The second term is a single-trace deformation of the CFT action, where $\mathcal{J}$ is the source (or coupling) for the gauge-invariant operator $\mathcal{O}[\phi]$ built from $\phi$. The factor $\sqrt{-\gamma}$ is the determinant’s square root of the induced conformal boundary metric \footnote{The boundary induced metric is $h_{ab}$, and $\gamma_{ab}$ its conformal representative, related by $h_{ab} = r_\infty^2 \gamma_{ab}$. The bulk line element is in equation \eqref{metric}.}.

In the saddle-point approximation, the field $\phi$ satisfies the saddle-point equation
\begin{equation}
\frac{\delta S_{\text{CFT}}[\phi^*]}{\delta \phi} + \int_{\Sigma} \sqrt{-\gamma}\, \mathcal{J} \frac{\delta \mathcal{O}[\phi^*]}{\delta \phi} = 0
\quad \Longrightarrow \quad \phi^* = \phi^*[\mathcal{J}]\,.
\end{equation}
Thus, $\phi^*$ in \eqref{bdry-theory-infty} denotes the boundary field satisfying the above saddle-point equation.

We conclude by emphasizing that the function $\mathcal{J}$ in the bulk Dirichlet boundary condition \eqref{D-bc} acts as a source or coupling in the boundary theory \eqref{bdry-theory-infty}. In other words, the bulk field’s boundary value serves as an external source for the corresponding CFT operator.
\section{Holography with arbitrary boundary condition}
As reviewed, the standard AdS/CFT formulation imposes Dirichlet boundary conditions on bulk fields. Here, we show how holography in the saddle-point approximation extends to arbitrary boundary conditions. The key claim is: modifying bulk boundary conditions corresponds to introducing multitrace deformations in the dual theory. This was first proposed in \cite{Witten:2001ua}, and we provide a simple derivation.

We start by asking how to modify bulk field boundary conditions. The answer is simple: add a boundary term to the bulk action. Specifically, we consider the modified action
\begin{equation}\label{bulk-action-W}
S^{\text{\tiny{W}}}_{\text{\tiny{bulk}}}[J;\mathcal{M}] = \int_{\mathcal{M}} \mathcal{L}_{\text{\tiny{bulk}}}^{\text{\tiny{W}}}[J]\, , \qquad \mathcal{L}_{\text{\tiny{bulk}}}^{\text{\tiny{W}}} = \mathcal{L}_{\text{\tiny{bulk}}}^{\text{\tiny{D}}} + \partial_{\mu}W^{\mu}\, ,
\end{equation}
Here, $\mathcal{L}{\text{\tiny{bulk}}}^{\text{\tiny{D}}}$ is the bulk Lagrangian with Dirichlet boundary conditions. Adding the total derivative $\partial{\mu}W^{\mu}$ defines a new action $\mathcal{L}_{\text{\tiny{bulk}}}^{\text{\tiny{W}}}$. While this does not change bulk equations, it modifies boundary conditions, called $W$-type. By choosing $W^{\mu}$ appropriately, we can move from Dirichlet to Neumann, mixed, and other boundary conditions. Examples appear in Section \ref{sec:comments}.

To understand the boundary interpretation of this modification, we start from the Dirichlet AdS/CFT correspondence in the saddle-point approximation \eqref{AdS/CFT} and include the same boundary term on both sides. This yields
\begin{equation}
\inbox{{S}^{\text{\tiny{W}}}_{\text{bdry}} [\mathcal{J};\Sigma] = S^{\text{\tiny{W}}}_{\text{bulk}} [\mathcal{J}; \mathcal{M}]\, ,}
\end{equation}
where the right-hand side is the modified bulk action compatible with $W$-type boundary condition, defined in \eqref{bulk-action-W}. On the left-hand side, the boundary action becomes
\begin{equation}\label{MTD-infty}
{S}^{\text{\tiny{W}}}_{\text{bdry}}[\mathcal{J};\Sigma] = {S}^{\text{\tiny{D}}}_{\text{bdry}}[\mathcal{J};\Sigma] + \int_{\Sigma} n_{\mu} W^{\mu}\, ,
\end{equation}
where, $n_\mu \mathrm{d}x^\mu = \mathrm{d}r$ is the unnormalized normal one-form to the AdS boundary $\Sigma$. The key point is that while the $W$-term appears as a codimension-one boundary term from the bulk perspective, it is codimension-zero from the boundary viewpoint, and thus acts as a genuine deformation of the boundary theory.

To complete the picture, we rewrite the $W$-term—originally in bulk variables—in terms of boundary ones, using the AdS/CFT dictionary for the required identifications
\begin{equation}\label{scaling-infty}
\mathcal{J}(x^a) = r_{\infty}^{d-\Delta} J(r_\infty, x^a)\, , \qquad \mathcal{O}(x^a) = r_{\infty}^{\Delta}\, O(r_\infty, x^a)\, ,
\end{equation}
The left-hand side gives the boundary source and operator, while the right-hand side shows the corresponding bulk fields near the boundary. Using this dictionary to rewrite the $W$-term in \eqref{MTD-infty} completes its interpretation as a multitrace deformation of the boundary action.
\section{Finite cutoff holography}
In the previous section, we generalized the AdS/CFT correspondence at the saddle-point level to accommodate arbitrary boundary conditions for bulk fields at the asymptotic AdS boundary. In this section, we take a further step by extending the correspondence to a finite radial position, thereby formulating a holographic framework at a finite cutoff.

\begin{figure}[t]
\centering
\begin{tikzpicture}[scale=1.3]

\node (v1) at (0.8,0) {};
\node (v4) at (0.8,-3) {};
\node (v5) at (-0.8,0) {};
\node (v8) at (-0.8,-3) {};

\begin{scope}
  \clip ($(v1)+(0,0)$) to[out=135,in=45, looseness=0.5] ($(v5)+(0,0)$)
        -- ($(v8)+(0,0)$) to[out=45,in=135, looseness=0.5] ($(v4)+(0,0)$)
        -- cycle;
  \fill[gray!20] (-0.8,0) -- (-0.8,-3) -- (0.8,-3) -- (0.8,0) -- cycle;
\end{scope}

\draw [darkred!60, very thick] (-0.8,0) -- (-0.8,-3);
\draw [darkred!60, very thick] (0.8,0) -- (0.8,-3);

\begin{scope}[fill opacity=0.8, very thick, darkred!60]
  \filldraw [fill=gray!30] (0.8,0) 
    to[out=135,in=45, looseness=0.5] (-0.8,0)
    to[out=315,in=225,looseness=0.5] (0.8,0);
\end{scope}
\begin{scope}[fill opacity=0.8, very thick, darkred!60]
  \filldraw [fill=gray!30] (0.8,-3) 
    to[out=135,in=45, looseness=0.5] (-0.8,-3)
    to[out=315,in=225,looseness=0.5] (0.8,-3);
\end{scope}

\draw [blue!60](0,0) ellipse (1.5 and 0.5);
\draw [blue!60] (0,-3) ellipse (1.5 and 0.5);
\draw [blue!60] (-1.5,0) -- (-1.5,-3);
\draw [blue!60] (1.5,0) -- (1.5,-3);

\fill (0,-1.5)  node [blue!50!black] {${\cal M}(r)$};
\fill (0.8,-2)+(0.3,0.5) node [darkred!60] {$\Sigma(r)$};
\fill (0.8,-3)+(0.7,0.8) node [right,blue!80] {$\Sigma$};
\fill (0, -3.8) node [black] { $\mathcal{M}$};
\end{tikzpicture}
\caption{ \justifying{\footnotesize{ Portion of an asymptotically AdS spacetime \cite{Adami:2025pqr}: the shaded region $\textcolor{blue!50!black}{\mathcal{M}(r)}$ is enclosed by a timelike surface $\textcolor{darkred!60}{\Sigma(r)}$. The full cylinder $\textcolor{black}{\cal{M}}$ denotes the global asymptotically AdS spacetime, with its asymptotic timelike boundary labeled $\textcolor{blue!80}{\Sigma}$.}}}
\label{fig:ADS-timelike}
\end{figure}
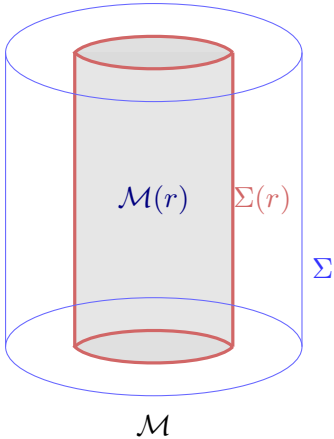

We begin by outlining the geometric setup. Let $\mathcal{M}$ be a $(d{+}1)$-dimensional asymptotically AdS spacetime with timelike boundary $\Sigma \equiv \partial \mathcal{M}$. To introduce a finite radial cutoff, we foliate the bulk by codimension-one hypersurfaces $\Sigma(r)$, defined as constant-$r$ slices with $r \in (r_\circ, \infty)$. Each $\Sigma(r)$ partitions spacetime, and we define the enclosed region as $\mathcal{M}(r)$, so that $\partial \mathcal{M}(r) = \Sigma(r)$ (see Fig.~\ref{fig:ADS-timelike}). Our aim is to formulate a holographic correspondence between $\mathcal{M}(r)$ and its boundary $\Sigma(r)$—effectively shifting the AdS boundary inward. In the limit $r \to \infty$, we recover the standard setup: $\mathcal{M}(r) \to \mathcal{M}$ and $\Sigma(r) \to \Sigma$.

We are now ready to develop finite cutoff holography in the saddle-point approximation. As reviewed, this limit reduces the standard correspondence to on-shell action equality. A similar relation is naturally expected at finite radial position
\begin{equation}\label{AdS/QFT}
    S^{\text{\tiny{D}}}_{\text{bulk}}[\mathcal{M}(r)] = {S}^{\text{\tiny{D}}}_{\text{bdry}}[\Sigma(r)]\, .
\end{equation}
This relation was explicitly derived in \cite{Parvizi:2025wsg} by starting from the standard AdS/CFT correspondence and applying covariant phase space techniques. The left-hand side of equation \eqref{AdS/QFT} has a clear interpretation—it is the on-shell bulk action evaluated in the region $\mathcal{M}(r)$
\begin{equation}
    S^{\text{\tiny{D}}}_{\text{bulk}}[\mathcal{M}(r)] = \int_{r_\circ}^{r} \mathrm{d}r \int_{\Sigma(r)} \mathcal{L}^{\text{\tiny{D}}}_{\text{bulk}}\, .
\end{equation}
Analogous to the gauge/gravity correspondence at the asymptotic boundary \eqref{D-bc}, the bulk fields are required to satisfy Dirichlet boundary conditions on the finite cutoff surface $\Sigma(r)$
\begin{equation}\label{D-bc-rc}
J(r,x^a)\Big|_{\Sigma(r)} = r^{\Delta - d} \mathcal{J}(r,x^a)\, , \qquad \delta \mathcal{J}(r,x^a) = 0\, .
\end{equation}

We now turn to the right-hand side of \eqref{AdS/QFT}: how is the boundary theory defined on the cutoff surface $\Sigma(r)$? As we will show, it corresponds to a specific multitrace deformation of the original theory on $\Sigma$. To derive the associated deformation flow equation, we consider the radial evolution of the on-shell bulk action.

Starting with a diffeomorphism-invariant theory on $\mathcal{M}(r)$, its variation under a general diffeomorphism gives
\begin{equation}\label{delta-xi-S-1}
    \delta_{\xi}S^{\text{\tiny{D}}}_{\text{bulk}}[\mathcal{M}(r)]=\int_{\Sigma(r)}\, n_{\mu}\, \xi^{\mu}\, \mathcal{L}^{\text{\tiny{D}}}_{\text{bulk}}\, .
\end{equation}
Choosing $\xi=\partial_{r}$, this becomes
\begin{equation}\label{delta-xi-S-1}
    \frac{\mathrm{d}}{\mathrm{d}r}S^{\text{\tiny{D}}}_{\text{bulk}}[\mathcal{M}(r)]=\int_{\Sigma(r)} \mathcal{L}^{\text{\tiny{D}}}_{\text{bulk}}\, .
\end{equation}
Using the finite-distance holographic relation \eqref{AdS/QFT}, we arrive at the flow equation
\begin{equation}\label{deform-flow-1}
  \inbox{  \frac{\mathrm{d}}{\mathrm{d}r} {S}^{\text{\tiny{D}}}_{\text{bdry}}[\Sigma(r)] = \int_{\Sigma(r)} \mathcal{L}^{\text{\tiny{D}}}_{\text{bulk}} \Big|_{\text{on-shell}}\, .}
\end{equation}
This equation is particularly noteworthy because both sides are defined on the same hypersurface $\Sigma(r)$. It captures how the boundary theory deforms as we move radially into the bulk, with $r$ serving as the deformation parameter. Being a first-order differential equation, its solution is uniquely determined once an initial condition is specified
\begin{equation}
    \lim_{r \to \infty} {S}^{\text{\tiny{D}}}_{\text{bdry}}[\Sigma(r)] = {S}^{\text{\tiny{D}}}_{\text{bdry}}[\Sigma]\, .
\end{equation}
Note that the right-hand side is determined using the standard gauge/gravity duality \eqref{AdS/CFT}. The remaining task is to rewrite the right-hand side of \eqref{deform-flow-1}, currently in bulk variables, in terms of boundary field theory data. Motivated by the scaling in \eqref{scaling-infty}, we introduce the following rescaled variables
\begin{equation}\label{scaling-rc}
\mathcal{J}(r,x^a) = r^{d-\Delta} J(r,x^a)\, , \qquad \mathcal{O}(r,x^a) = r^{\Delta}\, O(r,x^a)\, .
\end{equation}

We now derive a more explicit form of the boundary deformation. Recall that the variation of the on-shell bulk action under a diffeomorphism can also be written as
\begin{equation}\label{delta-xi-S-2}
    \begin{split}
    \delta_{\xi} S^{\text{\tiny{D}}}_{\text{bulk}}(r) = \int_{\Sigma(r)} \sqrt{-h} \, O \, \delta_{\xi} J\, ,
    \end{split}
\end{equation}
where $O$ is the radial canonical momentum conjugate to $J$. Setting $\xi = \partial_r$ gives radial flow
\begin{equation}\label{o-dr-J}
   \frac{\mathrm{d}}{\mathrm{d}r} {S}^{\text{\tiny{D}}}_{\text{bdry}}[\Sigma(r)] = \int_{\Sigma(r)} \sqrt{-h} \, O \, \partial_r J\, .
\end{equation}
Since $O$ is the radial momentum conjugate to $J$, we expect $\partial_r J \sim O$, which implies
\begin{equation}\label{deform-flow-2}
   \frac{\mathrm{d}}{\mathrm{d}r} {S}^{\text{\tiny{D}}}_{\text{bdry}}[\Sigma(r)] \sim \int_{\Sigma(r)} \sqrt{-h} \, O^2\, .
\end{equation}
This suggests that the boundary multitrace deformation induced by radial evolution is proportional to the square of the momentum $O^2$. In the case of pure gravity, where the momentum conjugate to the metric is the energy-momentum tensor, this structure directly resembles the T$\bar{\text{T}}$ deformation. As we will show below, this identification holds exactly.

So far, we have formulated finite cutoff holography with Dirichlet boundary conditions. We now extend this to arbitrary boundary conditions on $\Sigma(r)$ by adding $\int_{\Sigma(r)} \partial_{\mu} W^{\mu}$ to both sides of equation \eqref{deform-flow-1}. Then,
\begin{equation}\label{deform-flow-W-1}
  \inbox{  \frac{\mathrm{d}}{\mathrm{d}r} {S}^{\text{\tiny{W}}}_{\text{bdry}}[\Sigma(r)] = \int_{\Sigma(r)} \mathcal{L}^{\text{\tiny{W}}}_{\text{bulk}} \Big|_{\text{on-shell}}\, ,}
\end{equation}
where the right-hand side follows from \eqref{bulk-action-W}, and the left-hand side is given by
\begin{equation}
    {S}^{\text{\tiny{W}}}_{\text{bdry}}[\Sigma(r)] = {S}^{\text{\tiny{D}}}_{\text{bdry}}[\Sigma(r)] + W[\Sigma(r)]\, .
\end{equation}
This yields a generalized deformation flow equation at finite cutoff, now accommodating arbitrary boundary conditions. Its solution is fixed by the initial condition
\begin{equation}
    \lim_{r \to \infty} {S}^{\text{\tiny{W}}}_{\text{bdry}}[\Sigma(r)] = {S}^{\text{\tiny{W}}}_{\text{bdry}}[\Sigma] = {S}^{\text{\tiny{D}}}_{\text{bdry}}[\Sigma] + W[\Sigma]\, ,
\end{equation}
where ${S}^{\text{\tiny{D}}}_{\text{bdry}}[\Sigma]$  follows from standard AdS/CFT, and $W[\Sigma]$ encodes the chosen boundary condition. In the next section, we illustrate this with an explicit example.

{
We conclude this section with a comment on boundary Ward identities. In developing finite-cutoff holography, we first considered the variation of the bulk on-shell action under radial diffeomorphisms, which leads to the deformation flow equation \eqref{deform-flow-1}. We now turn to generic diffeomorphisms tangential to the boundary, satisfying $\xi_{||}^{\mu} n_{\mu}=0$. From \eqref{delta-xi-S-1}, we then conclude that, for tangential diffeomorphisms:
\begin{equation}\label{bdry-Ward-identity}
\delta_{\xi}S^{\text{\tiny{D}}}_{\text{bulk}}[\mathcal{M}(r)]=0 \quad \Longrightarrow \quad \delta_{\xi}S^{\text{\tiny{D}}}_{\text{bdry}}[\Sigma(r)]=0\, .
\end{equation}
This equation guarantees the invariance of the boundary theory under tangential diffeomorphisms. In contrast, normal diffeomorphisms induce boundary deformations, as described by \eqref{deform-flow-1}. This behavior is a direct consequence of introducing a boundary, which reduces the full bulk diffeomorphism invariance to tangential diffeomorphisms at the boundary. Equation \eqref{bdry-Ward-identity} then directly gives rise to the boundary Ward identities.}
\section{Example: Einstein's gravity}
Thus far, our construction has been presented in an abstract setting. To demonstrate its broad applicability, we now turn to a concrete example: pure Einstein gravity.

The Einstein–Hilbert action with a negative cosmological constant, compatible with Dirichlet boundary conditions on a $(d+1)$-dimensional region $\mathcal{M}(r)$, is 
\begin{equation}\label{action-GR}
   S^{\text{\tiny{D}}}_{\text{\tiny{bulk}}}[\mathcal{M}(r)] = \int_{\mathcal{M}(r)} \mathcal{L}_{\text{bulk}}^{\text{\tiny{D}}} = \frac{1}{2} \int_{\mathcal{M}(r)} \sqrt{-g} \left( \mathscr{R} + \frac{d(d-1)}{\ell^2} \right) + \int_{\Sigma(r)} \sqrt{-h} \, K \, ,
\end{equation}
where $\ell$ is the AdS radius. The boundary term is the Gibbons–Hawking–York (GHY) term,
ensuring a well-defined variational principle with Dirichlet conditions. We adopt the Fefferman–Graham gauge \cite{fefferman1985conformal}, where the bulk line element takes the form
\begin{equation} \label{metric}
    \mathrm{d}s^2 = \frac{\ell^2}{r^2} \, \mathrm{d}r^2 + h_{ab}(r, x) \, \mathrm{d}x^a \mathrm{d}x^b \, .
\end{equation}

The on-shell variation of the action yields
\begin{equation}\label{on-shell-var-action-D}
    \delta {S}^{\text{\tiny{D}}}_{\text{bulk}}[\mathcal{M}(r)] \Big|_{\text{on-shell}} 
    = -\frac{1}{2} \int_{\Sigma(r)} \sqrt{-h}\, T^{ab}\, \delta h_{ab}\, ,
\end{equation}
where $T^{ab}$ is the Brown-York energy-momentum tensor (BY-EMT) \cite{Brown:1992br}
\begin{equation}\label{BY-EMT}
    T^{ab} = K^{ab} - K h^{ab} \, , \qquad K_{ab} = \frac{r}{2\ell} \, \partial_r h_{ab}\, ,
\end{equation}
with $K_{ab}$ the extrinsic curvature of $\Sigma(r)$ and $K := h^{ab} K_{ab}$. The BY-EMT is the canonical conjugate to the induced metric $h_{ab}$ on $\Sigma(r)$. Equation~\eqref{on-shell-var-action-D} shows the action is stationary under variations $\delta h_{ab}$ vanishing on the boundary, as required by Dirichlet conditions.

Performing a $1+d$ decomposition of Einstein’s equations along the foliation by hypersurfaces $\Sigma(r)$ yields the system
\begin{subequations}\label{EoM-GR-mat-decompose}
    \begin{align}
       & R + \frac{d(d-1)}{\ell^2} + {\Ottbar}  = 0 \, ,\label{EoM-ss}\\
       & {\nabla}_bT^{b}_{a}=0\, , \label{EoM-sa}\\
       & \frac{\kappa r}{\ell}\partial_rT_{ab}-2\,T_{a}^{c}T_{bc} + \frac{TT_{ab}}{d-1} - {\Ottbar}  h_{ab}-{R}_{ab}=0\, , \label{EoM-ab} \\
       & r \partial_{r} h_{ab} = 2 \ell \left( T_{ab} - \frac{T}{d-1}h_{ab} \right)\, ,
    \end{align}
\end{subequations}
where the last line defines $T_{ab}$; $R$ and $R_{ab}$ are the Ricci scalar and tensor of the induced metric $h_{ab}$, and $\nabla_a$ is the covariant derivative compatible with $h_{ab}$. The quantity ${\Ottbar}$ is a covariant generalization of the T$\bar{\text{T}}$ operator to $d$ dimensions \cite{Hartman:2018tkw, Taylor:2018xcy}, defined as
\begin{equation}\label{TTbar-NR}
    {\Ottbar} = T^{ab} T_{ab} - \frac{1}{d-1} T^2 \, .
\end{equation}

As discussed in \eqref{deform-flow-2}, the deformation action is quadratic in the canonical momentum. For GR, starting from \eqref{o-dr-J}, we find \eqref{o-dr-J} for GR
\begin{equation}\label{radial-bulk-action-NR}
   \begin{split}
     \hspace{-.12 cm}  \frac{\mathrm{d}{}}{\mathrm{d}{}r} {S}^{\text{\tiny{D}}}_{\text{bulk}}[\mathcal{M}(r)] \Big|_{\text{on-shell}} & = -\frac{1}{2} \int_{\Sigma(r)} \sqrt{-h}\, T^{ab}\, \partial_{r} h_{ab} \\
       & = - \frac{\ell}{r}\int_{\Sigma(r)} \sqrt{-h}\, T^{ab}\, K_{ab} \\
       & = - \frac{\ell}{r}\int_{\Sigma(r)} \sqrt{-h}T^{ab} \left(T_{ab} - \frac{T}{d-1} h_{ab} \right) \\
       & = - \frac{\ell}{r} \int_{\Sigma(r)} \sqrt{-h}\, {\Ottbar}\, .
   \end{split}
\end{equation}
In the second and third lines, we used the definitions of the extrinsic curvature and the BY-EMT given in \eqref{BY-EMT}, while in the final line we employed the definition of the T$\bar{\text{T}}$ operator from \eqref{TTbar-NR}. Finally, invoking the finite-distance holographic relation \eqref{AdS/QFT}, we obtain
\begin{equation}\label{TTbar-deformation-flow-NR}
    \inbox{ r \frac{\mathrm{d}}{\mathrm{d}r} {S}^{\text{\tiny{D}}}_{\text{bdry}}[\Sigma(r)] = - \ell \int_{\Sigma(r)} \sqrt{-h}\, {\Ottbar}\, . }
\end{equation}
This foundational statement shows the T$\bar{\text{T}}$ deformation of the boundary theory admits a holographic interpretation as the radial flow of the AdS boundary into the bulk \cite{Hartman:2018tkw, Taylor:2018xcy, McGough:2016lol}. We have derived it explicitly within the saddle-point approximation in arbitrary dimensions.
\section{Comments on boundary deformation flow}\label{sec:comments}
In this section, we discuss key aspects of the boundary deformation flow equation \eqref{TTbar-deformation-flow-NR}.
 \paragraph{Gravity is induced by renormalization group flow.}
 One important point to note is that the deformation flow equation \eqref{TTbar-deformation-flow-NR} is subject to a constraint: it must satisfy the Hamiltonian constraint \eqref{EoM-ss}. By incorporating this condition, the flow equation can be recast in the following form
 \begin{equation}\label{GR-deformation-flow-NR}
    \inbox{ r \frac{\mathrm{d}}{\mathrm{d}r} {S}^{\text{\tiny{D}}}_{\text{bdry}}[\Sigma(r)] =  \ell \int_{\Sigma(r)} \sqrt{-h}\, \left( R + \frac{d(d-1)}{\ell^2} \right)\, . }
\end{equation}
This fundamental result shows that the T$\bar{\text{T}}$ deformation \eqref{TTbar-deformation-flow-NR} can be expressed geometrically, revealing a deep gravitational interpretation. Since the radial direction $r$ encodes the RG flow from UV (AdS boundary) to IR (bulk), equation \eqref{GR-deformation-flow-NR} implies gravity emerges dynamically along the RG flow of the originally non-gravitational boundary theory. In other words, gravity on the cutoff surface $\Sigma(r)$ is induced by the boundary deformation, highlighting the \textit{emergent} nature of gravitational dynamics in holography \cite{Adami:2025pqr}. 
\paragraph{Renormalized deformation flow equation.} 
In the previous section, we used the gravitational action \eqref{action-GR}, compatible with Dirichlet boundary conditions but yielding a divergent on-shell value as $r \to \infty$. To regularize, appropriate counterterms must be added \cite{Balasubramanian:1999re, Emparan:1999pm}
    \begin{equation}
        \mathcal{S}^{\text{\tiny{D}}}_{\text{\tiny{bulk}}}[\mathcal{M}(r)] =S^{\text{\tiny{D}}}_{\text{\tiny{bulk}}}[\mathcal{M}(r)] + \int_{\Sigma(r)} \sqrt{-h} \left[ \frac{d-1}{\ell} + \frac{\ell\, {R}}{2(d-2)} + \cdots\right]\, ,
    \end{equation}
    where the ellipsis represents additional curvature terms required in higher dimensions. Including these counterterms leads to the \textit{renormalized boundary flow equation}
    \begin{equation}\label{dr-S-D}
    \begin{split}
        r \frac{\mathrm{d}}{\mathrm{d}r} \mathcal{S}^{\text{\tiny{D}}}_{\text{bdry}}[\Sigma(r)]  = & - \ell \int_{\Sigma(r)} \sqrt{-h}\bigg[  \ttbar + \ell^{-1} \mathcal{T} + {\ell}{S}^{ab}\, \mathcal{T}_{ab} + \cdots \bigg]\, ,
    \end{split}
    \end{equation}
    where $\mathcal{T}_{ab}$ is the renormalized Brown-York energy-momentum tensor (rBY-EMT), given by
    \begin{equation}\label{rBY-EMT-def}
    \mathcal{T}^{ab} = T^{ab} + T_{\text{\tiny{ct}}}^{ab}\, , \qquad  T_{\text{\tiny{ct}}}^{ab} = \frac{d-1}{\ell} h^{ab} - \frac{\ell}{d-2} {G}^{ab}+ \cdots\, .
    \end{equation}
    In equation \eqref{dr-S-D}, $\ttbar = \mathcal{T}^{ab} \mathcal{T}_{ab} - \frac{1}{d-1} \mathcal{T}^2$ denotes the renormalized version of the T$\bar{\text{T}}$ operator introduced in \eqref{TTbar-NR}, and ${S}_{ab}$ is the Schouten tensor
     \begin{equation}
     \begin{split}
      &{S}_{ab} = {\frac{1}{d-2}}\left({R}_{ab}- \frac{{R}}{2(d-1)}h_{ab}\right)\, \quad d>2\, ,   \qquad \& \qquad S_{ab} = 0 \quad d=2\, .
     \end{split}
     \end{equation}
  \paragraph{Other boundary conditions.}
A natural way to generalize the gravitational variational principle is by adding boundary terms that modify the boundary conditions. A well-known example is the one-parameter family \cite{Liu:2024ymn}
    \begin{equation}
        {S}^{\text{\tiny{W}}}_{\text{bulk}}[\mathcal{M}(r)] =  {S}^{\text{\tiny{D}}}_{\text{bulk}}[\mathcal{M}(r)] + w\int_{\Sigma(r)} \sqrt{-h}\, K\, ,
    \end{equation}
where $K$ is the extrinsic curvature’s trace and $w$ characterizes the boundary condition. However, this class generally does not yield a finite on-shell action. Except for Dirichlet ($w=0$), the necessary counterterms to cancel divergences are unknown. Hence, we call these \textit{unrenormalized} boundary conditions.

To overcome this, we define a family of \textit{renormalized boundary conditions}
\cite{Parvizi:2025shq}
    \begin{equation}\label{RN-bc}
        \mathcal{S}^{\text{\tiny{W}}}_{\text{bulk}}[\mathcal{M}(r)] =  \mathcal{S}^{\text{\tiny{D}}}_{\text{bulk}}[\mathcal{M}(r)] + w\int_{\Sigma(r)} \sqrt{-h}\, \mathcal{T}\, ,
    \end{equation}
where $\mathcal{T}$ is the trace of the rBY-EMT \eqref{rBY-EMT-def}. Special $w$ values correspond to known boundary conditions: $0$ (Dirichlet), $1$ (Neumann), and $\frac{2}{d}$ (conformal). Other $w$ define a generalized class of renormalized conformal-type boundary conditions \cite{Liu:2024ymn}. Crucially, this formulation ensures a finite on-shell action and consistent variational principle for any $w$.

Starting from the action \eqref{RN-bc} with renormalized boundary conditions, we obtain the corresponding \textit{renormalized boundary flow equation}
     \begin{equation}
    \begin{split}
        r\frac{\mathrm{d}}{\mathrm{d}r} \mathcal{S}^{\text{\tiny{W}}}_{\text{bdry}}[\Sigma(r)] = - \frac{\ell}{2}\int_{\Sigma(r)} & \sqrt{-h}\,  \Bigg\{ (2+w-dw) \ttbar  + 2\left(1+w-\frac{dw}{2}\right) \frac{\mathcal{T}}{\ell} \\
        & +\, \ell (2-d w) {S}_{ab} \mathcal{T}^{ab} - \ell^2\, w\left({S}_{ab}S^{ab}-{S}^{2}\right)+\cdots  \Bigg\}\, .
    \end{split}
    \end{equation}
\paragraph{Boundary condition flow.}
In this letter, we showed that the T$\bar{\text{T}}$ deformation of the boundary theory leads to a finite cutoff formulation of holography. As a final remark, we highlight an equivalent perspective: the T$\bar{\text{T}}$ deformation can be seen as modifying bulk field boundary conditions. This was developed in the first part of the freelance holography program, showing that any multitrace boundary deformation corresponds to a bulk boundary condition change. Since T$\bar{\text{T}}$ is a specific multitrace deformation, it induces a flow from Dirichlet to a mixed boundary condition at the asymptotic boundary $\Sigma$. Determining this mixed condition involves solving the radial evolution of Einstein’s equations. {In three dimensions, this flow was explicitly worked out in \cite{Guica:2019nzm}, while its higher-dimensional generalization is currently being developed perturbatively.}
\section{Summary and outlook}
In this paper, we introduced the Freelance Holography program—an extension of the AdS/CFT correspondence within the saddle-point approximation—that generalizes holography to arbitrary timelike boundaries and boundary conditions. {In contrast to \cite{Parvizi:2025shq, Parvizi:2025wsg}, where the framework was developed within the full covariant phase space formalism, here we pursued a more direct route, streamlining the arguments and supplying alternative proofs so as to highlight the conceptual structure without unnecessary technical overhead.}

{We explored several key aspects of this framework. First, we emphasized the constrained nature of the boundary deformation flow equation, stressing that it must be analyzed in conjunction with both the Hamiltonian and momentum constraints. Second, we highlighted the natural emergence of boundary gravity along the RG flow in arbitrary spacetime dimensions, reinforcing the perspective that gravity is induced rather than fundamental. Third, we clarified the distinction between renormalized and unrenormalized boundary conditions, demonstrating their essential differences and their implications for holography at finite cutoff.}


A particularly striking implication of this approach is its reinterpretation of the renormalization group: by identifying radial flow in the bulk with deformation flow in the boundary theory, gravity emerges not as a fundamental force but as a collective effect induced by quantum field theoretic data. We believe this RG-based perspective may lead to a significant shift in how we understand Wilsonian renormalization in the presence of gravity.

{As a final point, we comment on how radiative modes—i.e., the bulk propagating modes or “news”—are encoded in the boundary theory. In our setup, the bulk gravitational degrees of freedom are fully captured by the boundary data. Working in the Fefferman–Graham gauge, these reduce to the conformal boundary metric and the holographic energy–momentum tensor, with the latter subject to both Hamiltonian and momentum constraints \cite{deHaro:2000xn}. Under Dirichlet boundary conditions, all radiative content is encoded in the boundary energy–momentum tensor. Being symmetric in $d$ dimensions, this tensor initially has $d(d+1)/2$ components; the trace constraint removes one, and the momentum constraints remove another $d$, leaving exactly $\tfrac{D(D-3)}{2}$ independent components, matching the expected number of bulk gravitons in $D=d+1$ dimensions. For more general boundary conditions, these propagating modes are distributed between the boundary metric and the stress tensor, but the total count of independent radiative modes remains unchanged, demonstrating the robustness of this identification across different boundary prescriptions.}

The Freelance Holography program is still in its early stages, with numerous open directions ranging from formal development to concrete applications. {We hope that the perspective developed here will serve as a foundation for further progress in understanding holography beyond the traditional AdS/CFT setup.}

\section*{Acknowledgment}
The author thanks H. Adami, M. Golshani, A. Parvizi, M.M. Sheikh-Jabbari, and M.H. Vahidinia for many insightful discussions, and M.R. Fadaei-Sadr for preparing the figure.

\addcontentsline{toc}{section}{References}
\bibliographystyle{fullsort.bst}
\bibliography{reference}

\end{document}